# Dispersion managed mode-locking in all-fiber polarization-maintaining Nd-doped laser at 920 nm

Aram A. Mkrtchyan, Mikhail. S. Mishevsky, Yuriy G. Gladush, Mikhail A. Melkumov, Aleksandr M. Khegai, Pavlos G. Lagoudakis, Albert G. Nasibulin

*Abstract*— **We present a comprehensive study of dispersion managed ultrashort-pulse generation in Nd-doped polarization maintaining all-fiber laser at a 920 nm wavelength. We implement a linear laser scheme with a chirped fiber Bragg grating (CFBG) serving as a semitransparent mirror and semiconductor saturable absorber mirror (SESAM) as a mode-locker and as a second non-transparent mirror. Complete 1064 nm emission filtering is ensured by 1064/1064 wavelength division multiplexer. The dispersion compensation by CFBG was sufficient to reach anomalous net dispersion in the shortest laser scheme allowing us to investigate mode-locking regimes in the -0.05 ps$^2$ ÷ 0.24 ps$^2$ net dispersion range by varying the length of the passive fiber. With this approach we demonstrate high order harmonic mode-locking at dispersion closer to zero and large energy span near-parabolic shape single pulse operation in anomalous dispersion regime. The investigation is supported by numerical simulations used to optimize the laser resonator design and to investigate intracavity pulse evolution.**

## I. INTRODUCTION

SINCE the development of the fiber mode-locked laser [1], ultrafast fiber lasers continue to captivate scientific community with their distinctive properties, including high stability, low cost, robustness, portability and etc. The most striking results are obtained in the spectral region between one and two microns, owning to selection of highly efficient active media in this region. For shorter wavelength Titanium:Sapphire laser sources are traditional and remain the "golden standard" for high power ultrashort pulse generation with a great output beam quality as well as for broad wavelength tuning, which make them universal tool for various purposes [2]. However, a number of significant drawbacks, such as bulkiness, sensitivity to mechanical perturbations and extremely high cost limits their practical applications outside laboratories.

Alternatively, since the first demonstration of a mode-locked laser at 920 nm in 1997 [3], significant progress has been made. A few techniques for the wavelength conversion were utilized for the ultrafast pulse generation in a sub-micron region:

Yb-doped fiber laser based optical parametric oscillator [4], [5], line-shift in a highly nonlinear fibers [6], [7], second harmonic generation from 1.8 to 1.9 µm ultrafast emission [8], fiber-optic Cherenkov radiation in photonic crystal fibers [9].

Mode-locking with direct generation using quasi-three-level $^4F_{3/2}$ - $^4I_{9/2}$ transition of Nd-doped fiber is another promising solution to cover the 900 nm band [3], [10]–[18]. This transition possesses much smaller gain compared to four-level transition at 1064 nm which needs to be filtered. Just a while ago all-polarization-maintaining (all-PM) Nd-doped fiber laser with free-space semiconductor saturable absorber mirror (SESAM) and bandpass filter was demonstrated emitting 8 ps pulses at the output compressed down to 1.3 ps with a double grating compressor [19]. In this and other works, all the publications so far have one of the following two disadvantages: bulk optical elements mostly grating pair systems for dispersion compensation and spectral filtering or non-PM fibers inside the laser cavity, adding instability to mechanical disturbances [3], [4], [8], [10], [11], [13]–[15], [17]–[22]. Recently, we published all-fiber PM mode-locked laser implementing nonlinear amplifying loop mirror (NALM) as a mode locking device [23]. Yet, it was in the long NALM scheme with dissipative soliton resonance mode-locking at a large positive dispersion, which limits its potential application due to low 6 MHz pulse repetition rate and pulse duration as long as 100 ps.

Here, for the first time we report a dispersion managed ultrashort-pulse generation in Nd-doped PM all-fiber laser at the 920 nm wavelength comprising chirped fiber Bragg grating (CFBG) for dispersion compensation and SESAM in a linear resonator scheme. We present a numerical and experimental investigation of self-starting mode-locked pulse generation regimes in -0.05 ps$^2$ ÷ 0.24 ps$^2$ net dispersion range by varying the passive fiber length. Moreover, the study of dispersion management allows us to produce harmonic mode-locking up to the 12$^{th}$ order with a 0.43 GHz pulse repetition rate at fairly close to zero net-dispersion for intracavity powers below the SESAM degradation threshold. The experimental part is supported by numerical simulation, used to optimize the cavity design and provide detailed information on intracavity pulse dynamics.

Acknowledgments; The reported study was funded by RFBR according to the research project № 20-32-90233. Y.G.G. and A.G.N. thank the Council on grants of the RF number НШ-1330.2022.1.3.

A. A. Mkrtchyan, M. S. Mishevsky, Y. G. Gladush and A. G. Nasibulin are with the Skolkovo Institute of Science and Technology, Nobel 3, Moscow 121205, Russia (e-mail: Aram.Mkrtchyan@skoltech.ru; Mikhail.Mishevsky@skoltech.ru; Y.Gladush@skoltech.ru; A.Nasibulin@skoltech.ru).

M. A. Melkumov, A. M. Khegai are with the Prokhorov General Physics Institute of the Russian Academy of Sciences, Dianov Fiber Optics Research Center, Moscow, 117942, Russia (e-mail: melkoumov@fo.gpi.ru; hegayam@gmail.com).

P. G. Lagoudakis is with the Skolkovo Institute of Science and Technology, Nobel 3, Moscow 121205 and Department of Physics and Astronomy, University of Southampton, Southampton SO17 1BJ, United Kingdom (e-mail: P.Lagoudakis@skoltech.ru).

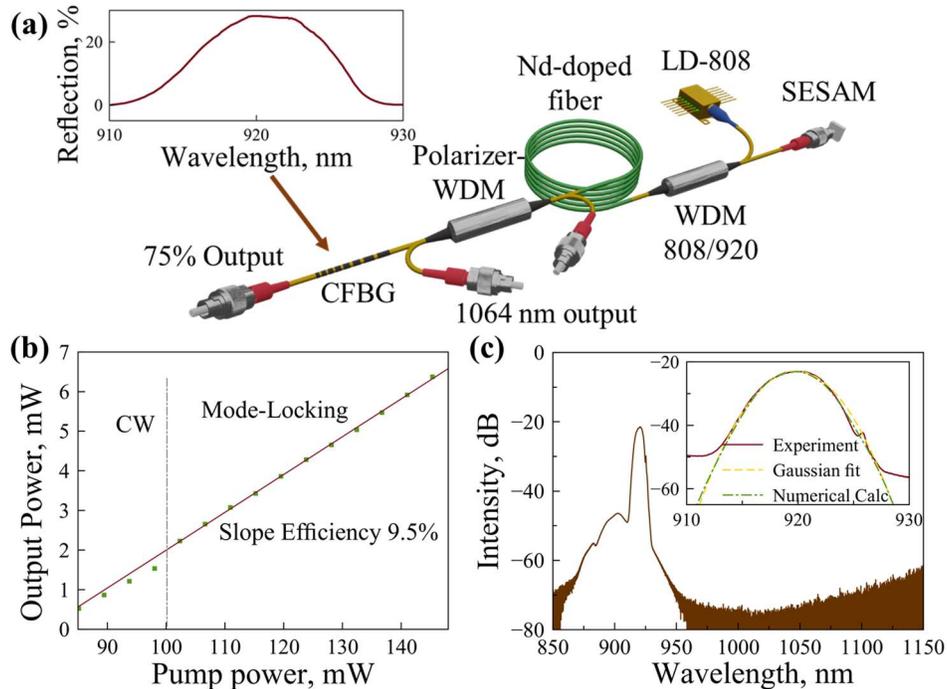

**Fig. 1.** a) Linear cavity mode-locked fiber laser scheme: CFBG – chirped fiber Bragg grating, WDM - wavelength division multiplexer, LD - laser diode, SESAM - semiconductor saturable absorber mirror, b) laser mode-locking threshold and slope efficiency, c) laser output pulse spectrum at 920 nm at the highest output power, limited by SESAM damage threshold of 10 mW. No sign of ASE at 1064 nm. Inset – experimental and numerical pulse spectra with Gaussian fit at anomalous net dispersion.

## II. EXPERIMENTAL SETUP

Linear laser scheme, used to obtain mode-locked generation, is shown in Fig. 1(a). Here, a 808 nm diode (LU0808M250 Lumics) pumps PM single-mode neodymium-doped silica fiber (CorActive ND 103-PM) with ~ 40 dB/m core absorption at 808 nm via 800 nm/920 nm wavelength division multiplexer (WDM). The length of the active fiber was optimized for the maximum efficiency, which were obtained at 1.3 m. From the one side cavity ends with chirped fiber Bragg grating (CFBG, TeraXion) with -0.19 ps² group delay dispersion, used to compensate large normal dispersion of the fibers inside the cavity. It provides 25% of reflection, taking 75% out. The CFBG also serves as a bandpass filter with 920 nm central wavelength and Gaussian profile with 10 nm bandwidth at 3 dB (see Fig. 1(a) inset), thus defining the laser emission wavelength. To obtain mode-locking we implemented SESAM (BATOP GmbH) mounted on the connector ferrule on another end of the linear cavity. SESAM is specified for 920 nm with 15% modulation depth, 0.5 ps relaxation time, 30 μJ/cm² saturation fluence and degradation threshold of 10 mW average power. The dominating emission at 1064 nm is suppressed with WDM-1064/920 with one blocked axis so that laser emission propagates only along slow axis. Total cavity length with full roundtrip in the case of shortest laser scheme is 4.0 m, which, in combination with CFBG, give -0.05 ps² of anomalous net cavity dispersion. Further dispersion adjustment we realize by adding passive fiber with $\beta_2 = +35$ ps²/km normal dispersion in between SESAM and 800/920 WDM. Stable self-starting pulse generation is achieved starting from 100 mW pump power with 9.5% slope efficiency, while for lower pump powers laser produce continuous wave emission with slightly less efficiency, which can be explained by higher losses of the SESAM at CW than in mode-lock regime (see Fig. 1(b)). The lasing pulse spectrum at the highest output power in large scale is shown in Fig. 1(c) without any signs of parasitic ASE emission of Nd-fiber at 1064 nm. The emitted pulse spectrum at anomalous net dispersion has a Gaussian profile (see Fig. 1(c) inset) proved by numerical simulation, which will be discussed later.

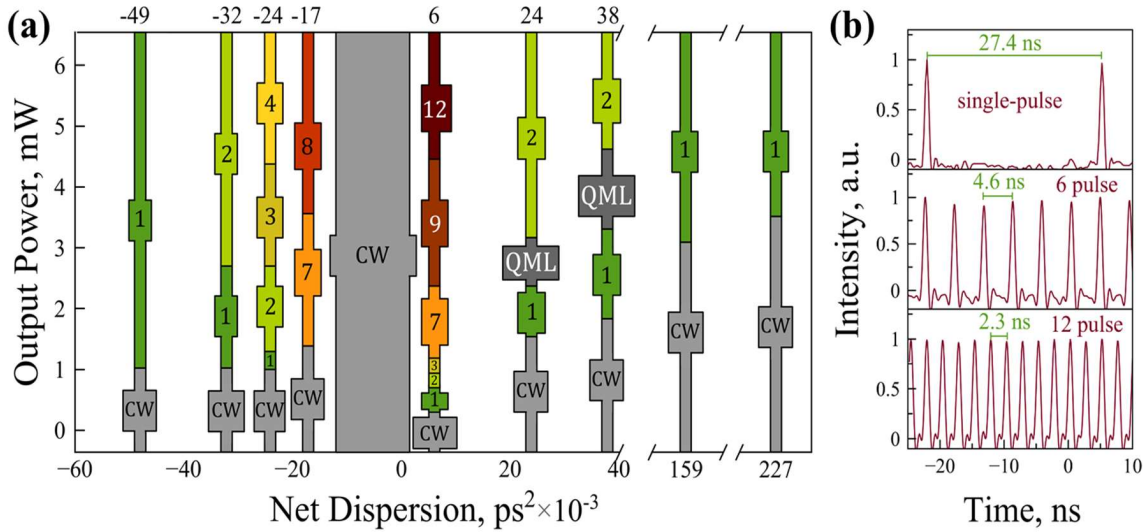

**Fig. 2.** a) Map of obtained laser generation regimes for various net dispersions and output powers. Numbers correspond to the order of harmonic mode-locking, CW denotes continuous wave generation and QML states for Q-switched mode locked transition regime. b) Harmonic mode-lock pulse trains of the 1st, 6th and 12th orders at near-zero net dispersion.

The characterization of the pulsed generation we perform with the following instruments: Yokogawa AQ6373 OSA, ThorLabs InGaAs biased detector DET08CFC and Tektronix MDO3102 1 GHz oscilloscope, Femtochrome FR-103WS autocorrelator, ThorLabs S132C photodiode power sensor.

## III. RESULTS AND DISCUSSIONS

When net dispersion approaches zero, its relatively small variation can lead to significant change in pulse generation regime [15], [24]–[26]. In this experiment we varied net dispersion of the cavity in -0.05 ps² ∼ 0.24 ps² range and analyzed on the map the pulsed generation regimes depending on the intracavity power, and corresponding output power, which is shown in Fig. 2(a). The intracavity power is limited by the SESAM degradation threshold of 10 mW average power, specified by the manufacturer, so we set the limit to 7 mW on the output. At closer to zero net anomalous dispersion, pulse generation threshold and soliton splitting threshold are decreasing in line with soliton area theorem. It results in the reduction of maximum achievable single pulse energy and promotion of multiple pulse generation [27], [28]. In case of zero net dispersion pulse generation is not observed. Typically, significantly higher modulation depth of saturable absorber is required at this point for stable mode locking [29], as pulse energy gets too small to induce sufficient absorber saturation. According to our numerical simulation we would require at least 30% of modulation depth while having 15% in the experiment. At slightly positive dispersion around +0.006 ps², the system tends to generate in harmonic mode locking regime with minimum pulse energy. We were able to reach up to 12th harmonic with 0.43 GHz pulse repetition rate, 14 pJ energy and 5.5 ps width, limited by SESAM damage threshold, see Fig. 2(b). Further increase towards positive net dispersion expands the intracavity power range that supports single pulse generation. At the same time, jumping from single to double pulse happens throughout the Q-switched mode-locking

(QML). For the long cavities, starting from 0.15 ps² net dispersion, the pulses demonstrate behavior more typical for dissipative solitons, with flat-top spectrum, highly chirped pulses and large single pulse operation range. At dispersion larger than 0.24 ps² we were not able to observe mode locking at available intracavity power of the laser.

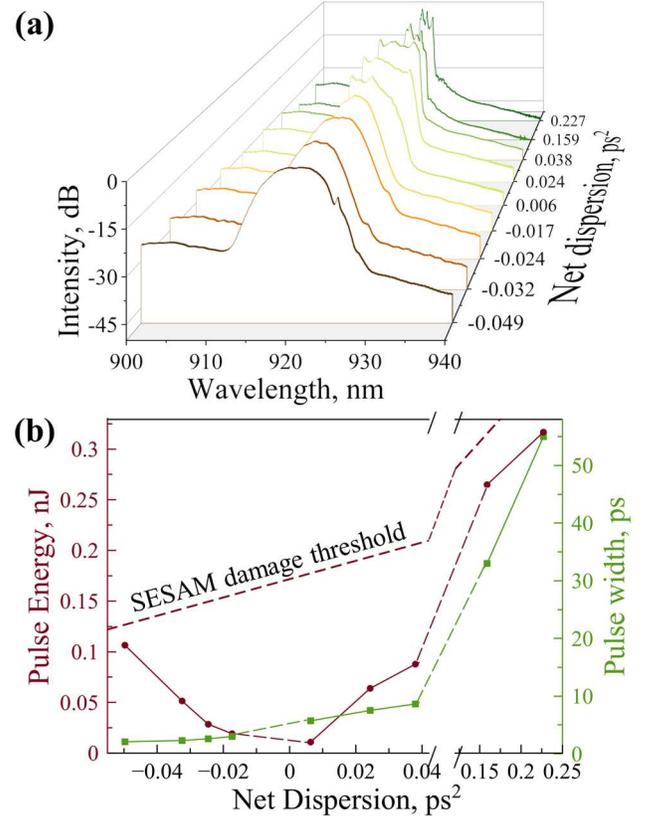

**Fig. 3.** a) Output pulse spectra for different net cavity dispersion. b) Pulse energy (brown) and pulse width (green) at different net dispersion. Brown dashed line shows SESAM damage threshold corresponding to the fixed average power calculated from saturation fluence inclined due to change of the cavity length. Lines connecting the point are presented for the eyeguiding.

The spectral measurements show dramatic change of the pulse spectra for different net cavity dispersions (Fig. 3(a)). For long cavities the total fiber normal dispersion few times exceeds the anomalous dispersion of CFBG and the net dispersion remains large and positive. Generated pulse leaves the cavity propagating through the CFBG without dispersion compensation at the end of roundtrip and its spectrum obtains the shape, typical for dissipative soliton pulse. The increase of the cavity length causes the pulse broadening and reducing the peak power, which leads to narrowing of the pulse spectrum due to decrease of self-phase modulation [19]. For short cavities with anomalous net dispersion pulse obtain spectra close to Gaussian shape on the output (see Fig. 1(c) inset), whose intracavity evolution will be discussed further. All the listed pulse characteristics are in good agreement with dispersion managed soliton behavior [19], [25], [26], [30], [31], with exception of the largest $0.15 \div 0.24$ ps$^2$ normal dispersion case, where dissipative soliton pulse behavior is observed.

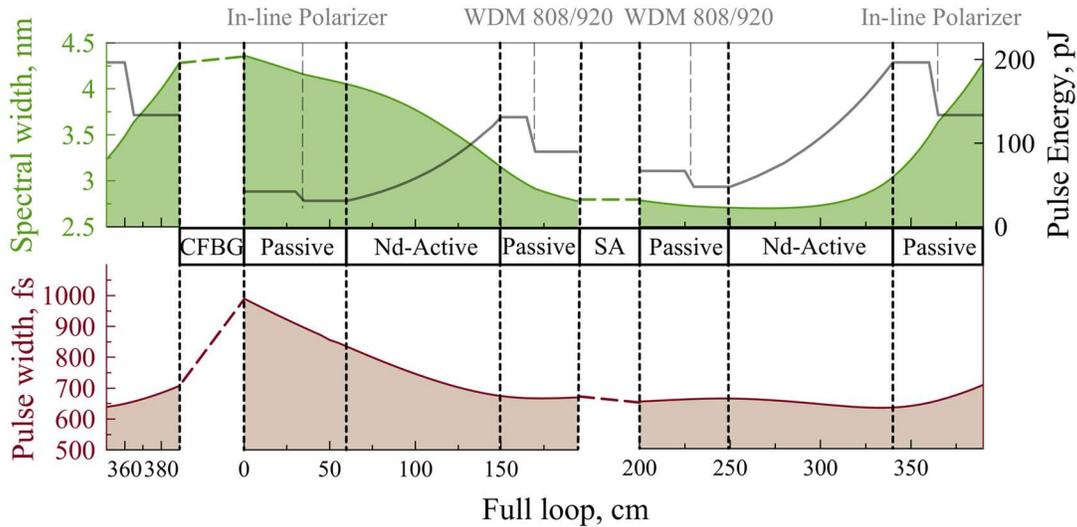

**Fig. 4.** Pulse energy, spectral and temporal widths dynamics inside the laser cavity.

The pulse energy and pulse duration as a function of net cavity dispersion are summarized in Fig. 3 (b). As discussed earlier, the smallest pulse energy of 14 pJ is observed near zero dispersion and increased as we move away from this point. For the largest normal and anomalous dispersions pulse energy is limited by the SESAM damage threshold line (Fig. 3(b), brown dashed line) corresponding to the fixed average power calculated from saturation fluence inclined due to change of cavity length. In contrast, the pulse duration demonstrates monotonic growth when moving from negative to positive dispersion (see Fig. 3(b), green curve). Propagating along the optical fibers with normal dispersion the pulse accumulates positive chirp. In the output of the laser CFBG doesn't affect the pulse chirp, whereas in reflection CFBG compensates the dispersion and can even change the sign of the pulse chirp. Thus, the output pulse always has positive chirp, which is proportional to the pulse duration and becomes larger with increase of the cavity length. Obtained pulses reaches 1.8 ps width at the shortest cavity scheme (-0.049 ps$^2$) and exceeds 50 ps in case of 11.6 m roundtrip length (+0.227 ps$^2$).

Finally, we investigate pulse dynamics and its intracavity behavior for the most important case of -0.05 ps$^2$ net dispersion, where shortest pulse and largest peak power is observed with 51 MHz repetition rate and 130 pJ pulse energy. For this purpose, we made numerical simulations with the Ginzburg-Landau equation for a complex amplitude based on split-step Fourier method [23], [32], [33]. Dispersion and Kerr nonlinearity of all fibers is taken +35 ps$^2$/km and 0.01 W$^{-1}$m$^{-1}$, respectively. The SESAM saturable absorber is modeled using the standard rate equation [34]. Propagation through the CFBG is calculated by multiplying the complex electric field to the reflection spectrum and dispersion coefficient in Fourier domain. Nd-doped fiber amplification modelled by Ginzburg-Landau equation with saturable gain with 50 nm parabolic bandwidth having maximum at 905 nm and 20 mW saturation power, which is estimated from Nd ASE emission at the 900 nm band [23].

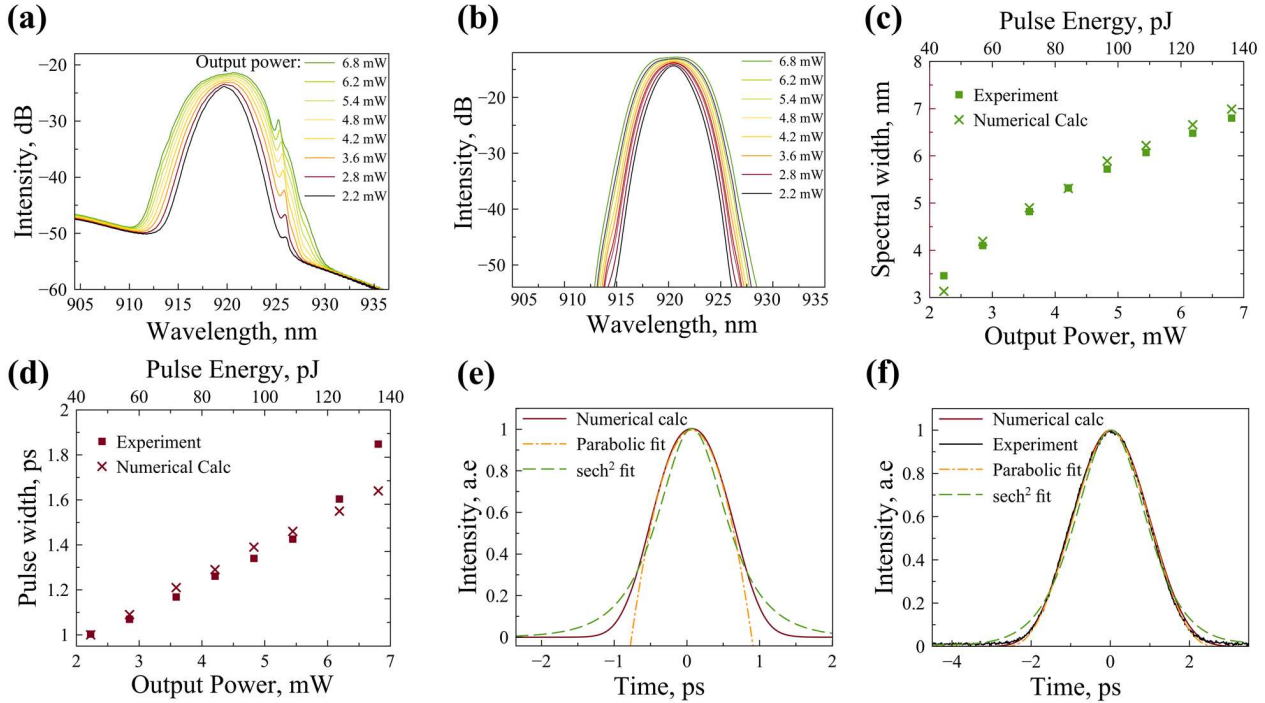

**Fig. 5.** a) Experimental and b) calculated pulse spectrum for different pump powers at anomalous net dispersion. Comparison of the experimental and calculated spectral c) and temporal d) widths of the pulse e) Simulated pulse shape on the output of the laser and its *sech²* and parabolic fit. f) Experimental and simulated autocorrelation traces of the pulse and its *sech²* and parabolic fits.

Despite pronounces anomalous dispersion, the pulsed regime shows no signs of conservative soliton behavior and demonstrates properties typical for the dispersion managed soliton.

According to calculations the pulse duration constantly changes along the resonator (Fig. 4). After the reflection from the CFBG with anomalous dispersion, the pulse changes the sign of the chirp from positive to negative, and then when propagating along the fiber with positive dispersion changes the sign of the chirp again. At this point the pulse reaches the minimum width (bottom panel) inside the resonator ~600 fs and also minimum spectral width (upper panel), which is also breathing within the cavity. This behavior is inherent for dispersion-managed soliton, which changes the sign of chirp twice as it propagates through the cavity [25], [26], [30], [31]. The corresponding pulse energy is shown in an upper panel with a black line. It is seen that the pulse is extracted near its maximum energy. Fig. 5(a, b) shows the experimental pulse spectra and the corresponding simulations for the different pump powers, which demonstrate good agreement. The increase in the pump power does not lead to a single pulse regime breaking and is accompanied by the growth of the spectral and temporal widths of the pulse (see Fig. 5(c, d)). This behavior is in striking contrast with expected conservative soliton properties and might indicate on the tendency towards self-similar parabolic pulse regime, which promotes single pulse operation [35]. Typically, similaritons are observed in the normal dispersion cavities with large gain per roundtrip. Never the less, recently similariton lasers in anomalous net dispersion ware demonstrated with a value of dispersion even higher than in present work [37], [38]. Our cavity has all required ingredients for the similariton generation: normal dispersion gain fiber with high self-modulation coefficient that creates the attractor and spectral filtering with anomalous dispersion segment [36]. Fig. 5(e) shows numerical simulation of pulse shape on the output of the laser and it's fitting with $sech^2$ and parabolic functions. The parabolic function provides much better agreement with simulation and experiment rather than $sech^2$, which autocorrelation traces are compared in Fig. 5(f). We emphasize that the transition from dispersion managed soliton to similariton depending on cavity parameters is smooth and it may be difficult to distinguish between these two regimes under certain conditions. We expect that the increase of the gain fiber would to parabolic regime more pronounced, however, the low damage threshold of SESAM limits available intracavity power. Nevertheless, parabolic regime is beneficial for further pulse amplification and opens a route for high energy pulse generation at 920 nm in all-fiber configuration.

## IV. CONCLUSION

In conclusion, we presented a dispersion managed ultrashort pulse generation at the 920 nm wavelength in the Nd-doped polarization maintaining all-fiber laser. Linear laser scheme is developed with chirped fiber Bragg grating as a semi-transparent output mirror and SESAM as a second fully reflecting mirror. The dispersion compensation by the chirped fiber Bragg grating allowed us to demonstrate various pulse regimes depending on the net dispersion. We obtain either parabolic-shape 130 pJ pulse, 2 ps pulse duration at 51 MHz repetition rate at anomalous net dispersion, or 14 pJ pulse, 5.5 ps width at 430 MHz repetition rate in a harmonic mode-locked regime at near zero dispersion. The damage threshold of SESAM sets the limits for the pulse energy and maximum repetition rate in the harmonic mode-locked regime, which can be expanded by using an appropriate saturable absorber or further amplification.

**Aram A. Mkrtchyan** was born in Yerevan, Armenia on Nov. 3, 1994. He received the B.S and M.S. degrees in applied mathematics and physics from Moscow Institute of Physics and Technology, Moscow, Russia and Skolkovo Institute of Science and Technology, Moscow, Russia. He received Ph.D. degree in Optics at Skolkovo Institute of Science and Technology. His research interest includes the development of ultrafast all-fiber lasers at wide spectral ranges including submicron wavelengths. He is a winner of RFBR grant for Ph.D. student, where is working on development of optical devices, with controllable nonlinearity based on carbon nanomaterials for applications in nonlinear Optics: ultrafast generation with controllable pulses regimes, four-wave mixing.

**Mikhail S. Mishevsky** was born in Biysk, Russia on Apr. 24, 1997. He received the B.S. degree in applied mathematics and physics from Novosibirsk State, Novosibirsk, Russia and the M.S. degree from Skolkovo Institute of Science and Technology, Moscow, Russia. His research interests include: ultrafast lasers with different pulses generation regimes, development of ultrafast all-fiber lasers at submicron wavelength range.

**Yuriy G. Gladush** was born in Moscow, Russia, on Aug. 16, 1983. He received the Ph.D. degree in theoretical investigation of nonlinear wave phenomena in Bose-Einstein condensates and optics from Institute for Spectroscopy Russian Academy of Sciences, Troitsk, Moscow.
Gladush is an Assistant Professor in Laboratory of Nanomaterials of Skolkovo Institute of Science and Technology. His main research area is optics and laser physics. In Nanomaterials lab he is responsible for projects related to optical applications of carbon nanotubes and other nanomaterials including fiber laser ultrashort pulse generation and SWCNT bolometer development. Prior to Skoltech prof. Gladush was working in Institute of spectroscopy where his research was dedicated to resonance energy transfer in organic/inorganic semiconductor hybrid structures.

**Mikhail A. Melkumov** was born in Moscow, Russia, on Nov. 21, 1978. He received the Graduate degree from the Physics Department, Moscow State University, Moscow, Russia, in 2001, and the Ph.D. degree from the Fiber Optics Research Center, Russian Academy of Sciences, Moscow, in 2006. Currently, he is the Head of the Fiber Lasers and Amplifiers Laboratory, Prokhorov General Physics Institute of the Russian Academy of Sciences, Dianov Fiber Optics Research Center. His research interests include Raman and rear-earth-doped fiber lasers, bidoped fiber lasers and amplifiers, and spectroscopy of active centers in silica-based glasses and fibers.

**Aleksandr M. Khegai** was born in Sovietabad, Andijan region, Russia, on Apr. 23, 1991. He received the B.S. and M.S. degrees in physics from Volgograd State Technical University, Volgograd, Russia, in 2014. He received Ph.D. degree in laser physics in Prokhorov General Physics Institute of the Russian Academy of Sciences, Dianov Fiber Optics Research Center. His research interests include spectroscopy of the rare-earth- and bismuth-doped active fibers, the development of continuous wave and ultrafast fiber lasers and amplifiers.

**Pavlos G. Lagoudakis** graduate of the University of Athens, Greece. Pavlos received his PhD degree in Physics from the University of Southampton, UK in 2003 and conducted his postdoctoral research on optoelectronic properties of organic semiconductors at the Ludwig Maximilians University of Munich, Germany. In 2006, he returned to Southampton as Lecturer at the department of Physics and Astronomy, where he combined his expertise in inorganic and organic semiconductors and set up a new experimental activity on Hybrid Photonics.
In 2008, Lagoudakis was appointed to a personal chair at the University of Southampton. From 2011 to 2014, Pavlos chaired the University's Nanoscience Research Strategy Group, an interdisciplinary research group of ~100 academics across physics, chemistry, maths, engineering, biology and medicine. From 2013 to 2019, Pavlos was the Director for Research at the department of Physics and Astronomy at the University of Southampton. At Skoltech, prof. Lagoudakis designed, setup and now consults the Hybrid Photonics Group with a focus on hybrid LEDs, PVs and spinotronics.

**Albert G. Nasibulin** was born in Novokuznetsk, Kemerovo Region, Russia on March 23, 1972. He got his PhD in Physical Chemistry (1996) at Kemerovo State University (Russia) and Doctor of Science (Habilitation, 2011) at Saint-Petersburg Technical State University (Russia). He is a Professor at Skolkovo Institute of Science and Technology and an Adjunct Professor at the Department of Chemistry and Materials Science of Aalto University School of Chemical Engineering. He held a post of the Academy Research Fellow in Academy of Finland from 2006 to 2011. Since 2018 he is a Professor of the Russian Academy of Sciences.
Prof. Nasibulin has specialized in the aerosol synthesis of nanomaterials (nanoparticles, carbon nanotubes and tetrapods), investigations of their growth mechanisms and their applications. He has a successful background in an academic research with about 300 peer-reviewed scientific publications and 40 patents. He is a co-founder of three companies: Canatu Ltd. (spin-off from Helsinki University of Technology, Finland) and CryptoChemistry and Novaprint (spin-offs from Skolkovo Institute of Science and Technology, Moscow, Russia).